\documentclass[showpacs,epsf,twocolumn]{revtex4-1}
\usepackage{float}
\usepackage{amsmath}
\usepackage{color,graphicx}
\graphicspath{ {images/} }
\usepackage{hyperref}
\begin{document}

\title{Probing lattice dynamics and electron-phonon coupling in topological nodal-line semimetal ZrSiS}

\author{Ratnadwip Singha$^{1}$, Sudeshna Samanta$^{2}$, Swastika Chatterjee$^{3}$, Arnab Pariari$^{1}$, Dipanwita Majumdar$^{1}$, Biswarup Satpati$^{1}$, Lin Wang$^{2\sharp}$, Achintya Singha$^{4\dagger}$, Prabhat Mandal$^{1}$}

\email{prabhat.mandal@saha.ac.in}

\email{$^{\dagger}$achintya@jcbose.ac.in}

\email{$^{\sharp}$wanglin@hpstar.ac.cn}

\affiliation{$^{1}$Saha Institute of Nuclear Physics, HBNI, 1/AF Bidhannagar, Kolkata 700 064, India}

\affiliation{$^{2}$Center for High Pressure Science and Technology Advanced Research, Shanghai, China}

\affiliation{$^{3}$Department of Physics, Indian Institute of Technology Kharagpur 721 302, India}

\affiliation{$^{4}$Department of Physics, Bose Institute, 93/1, Acharya Prafulla Chandra Road, Kolkata 700 009, India}

\date{\today}

\begin{abstract}
Topological materials provide an exclusive platform to study the dynamics of relativistic particles in table-top experiments and offer the possibility of wide-scale technological applications. ZrSiS is a newly discovered topological nodal-line semimetal and has drawn enormous interests. In this report, we have investigated the lattice dynamics and electron-phonon interaction in single crystalline ZrSiS using Raman spectroscopy. Polarization and angle resolved measurements have been performed and the results have been analyzed using crystal symmetries and theoretically calculated atomic vibrational patterns along with phonon dispersion spectra. Wavelength and temperature dependent measurements show the complex interplay of electron and phonon degrees of freedom, resulting in resonant phonon and quasielastic electron scatterings through inter-band transitions. Our high-pressure Raman studies reveal vibrational anomalies, which were further investigated from the high-pressure synchrotron x-ray diffraction (HPXRD) spectra. From HPXRD, we have clearly identified pressure-induced structural transitions and coexistence of multiple phases, which also indicate possible electronic topological transitions in ZrSiS. The present study not only provides the fundamental information on the phonon subsystem, but also sheds some light in understanding the topological nodal-line phase in ZrSiS and other iso-structural systems.
\end{abstract}

\maketitle

\section{Introduction}

Over the last few years, condensed matter physics community has
witnessed enormous advancement in band theory, which
leads to the discovery of topological insulators (TIs)
\cite{Xia:2009,Chen:2009} and topological semimetals (TSMs)
\cite{Liu:2014,Liu2:2014,Weng:2015,Lv:2015,Xu:2015} and has
triggered intense investigations to find materials with exotic
topological phases of matter. While these systems are fascinating
due to the intriguing physics of relativistic particles,
they also appear to be promising candidates for technological
applications \cite{Pesin:2012,Zhang:2012}. Topological materials are
characterized by unique surface and bulk states with topologically
distinct electronic band structures. A TI possesses finite band gap
in the bulk state, whereas its surface state hosts linear band
crossings. On the other hand, in TSMs, bulk conduction and valence
bands cross each other at either fourfold (Dirac node)
\cite{Liu:2014,Liu2:2014} or twofold degenerate (Weyl node) points
\cite{Weng:2015,Lv:2015,Xu:2015}, depending on the inherent
symmetries of the system. Near these nodes, the electronic bands
obey linear dispersion relation and provide the exclusive platform
to study the dynamics of long-sought Dirac/Weyl fermions. Presently,
the primary objective in this field is not only to find new
materials with such unique band structure but also to look beyond
the Dirac/Weyl-type excitations.

From band structure calculations, recently, the monolayers of
\textit{WHM}(\textit{W}=Zr, Hf; \textit{H}=Si, Ge, Sn; \textit{M}=O,
S, Se, Te) family of materials have been proposed to be
two-dimensional TIs \cite{Xu3:2015}. These isostructural compounds have identical electronic band structure. Although
TI bands have been observed on the surface of bulk ZrSnTe crystal
\cite{Lou:2016}, ARPES and transport experiments have revealed
topological nodal-line semimetal phase in several other members of
this family \cite{Schoop:2016,Ratnadwip:2017,Ali:2016,Hu:2016}. In
these systems, the bulk conduction band and valence band cross along an
one-dimensional line in \textbf{k}-space instead of discrete points,
thus making them even more interesting from the perspective of
fundamental physics. ZrSiS is the starting member of this group and
hosts multiple linear band crossings at different energy values of
bulk band structure, which form a nodal-line \cite{Schoop:2016}.
Remarkably, in ZrSiS, the linear band crossing persists up to an
energy range $\sim$2 eV (largest reported so far) and hence, can
provide the robust topological system desired for industrial
applications. Moreover, the quasi two-dimensional Fermi surface,
extremely large and anisotropic magnetoresistance and multi-band
quantum oscillations in ZrSiS, have already been the focus of a
number of studies \cite{Ratnadwip:2017,Ali:2016}. These reports
inferred that the layered structure with square Si atom sublattice,
essentially controls the unique topological properties in ZrSiS and other isostructural systems. In fact, it is well
established that the structural symmetries play a key role in
protecting the accidental degeneracies and non-trivial topological
electronic bands in a material. On the other hand, the system can be
driven to electronic topological transitions by symmetry-breaking.
Raman scattering is a well-known nondestructive powerful tool for
studying the structural symmetries of a material and to probe the
impact of phonon on electronic energy bands \cite{Caramazza:2016,Marini:2012,Postorino:2002}. Together with
low-temperature and high-pressure techniques, this method enables us
to draw an in-depth picture of the phonon dynamics, electron-phonon
interaction and possible structural and topological phase
transitions. These information are crucial for understanding the
structural, thermal, and electronic properties, and may help to
comprehend the mechanism of the observed topological nodal-line
semimetal phase in ZrSiS as well as in other members of the
\textit{WHM} family.

In this report, we have done Raman spectroscopy measurements on
single crystalline ZrSiS. The Raman modes have been probed in the
\textit{basal plane} (crystallographic \textit{c}-axis is parallel
to the wave vector $k_{i}$ of the incident laser) and \textit{edge
plane} ($k_{i}$ is perpendicular to the \textit{c}-axis)
configurations. All the Raman active phonon modes at the Brillouin
zone center and their vibrational patterns have been identified from the first-principles
calculations as well as crystal symmetry analysis. The polarization and crystal-angle-resolved
measurements have been performed. From the wavelength and
temperature dependent Raman studies, we have identified the complex interplay of phonons with electronic degree of freedom.
To further explore the phonon dynamics, Raman scattering experiments
have been done under high-pressure. Pressure is an effective and
clean way to tune the lattice as well as electronic states and is
expected to have a significant impact on the layered structure of
ZrSiS. Apart from the softening of the phonon modes, we have observed pressure-induced vibrational anomalies in the Raman spectra. To explain the observed results, high-pressure synchrotron x-ray diffraction (HPXRD) measurements have been performed. HPXRD spectra reveal structural phase transitions along with the coexistence of multiple phases, which may also be accompanied by the electronic topological transition in ZrSiS.\\

\section{Experimental and computational details}

The single crystals of ZrSiS were synthesized in iodine vapor transport method \cite{Haneveld:1964}. The polycrystalline powder was prepared from solid state reaction of elemental Zr (Alfa Aesar 99.9\%), Si (Strem Chem. 99.999\%) and S (Alfa Aesar 99.9995\%) in two steps. Si and S (in stoichiometric ratio) were sealed in an evacuated quartz tube and heated at 1000$^{\circ}$C for 48 h. The resultant powder was mixed with Zr and again sealed in a quartz tube under vacuum and heated at 1100$^{\circ}$C for 48 h. The obtained polycrystalline powder together with iodine, were finally sealed in another quartz tube under dynamic vacuum. This tube was placed in a gradient furnace for 72 h. During this period, the polycrystalline powder was kept at 1100$^{\circ}$C, whereas the temperature at other end of the tube was maintained at 1000$^{\circ}$C. Shiny, plate-like single crystals were obtained at the cooler end. As grown crystals were characterized by high-resolution transmission electron microscopy (HRTEM) using an FEI, TECNAI G$^{2}$ F30, S-TWIN microscope operating at 300 kV and equipped with a GATAN Orius SC1000B CCD camera. The elemental compositions were checked by energy dispersive X-ray spectroscopy in the same microscope.

The ambient pressure and low-temperature Raman spectra were collected with a LABRAM HR 800 systems, which is equipped with a spectrometer of 80 cm focal length, 1800 gr/mm
grating and a Peltier cooled CCD. Lasers of wavelength 488, 633 and 785 nm were used to excite the sample. A 100X objective with NA 0.9 was used to focus the laser beam on the crystal.

A symmetric DAC with culet size 300 $\mu$m was prepared to generate
high pressure up to 57 GPa (Fig. 7 in Appendix). A stainless steel or rhenium gaskets were preindented with 36 $\mu$m
thickness, followed by laser-drilling the central part to make a 170
$\mu$m-diameter hole. We made sure that the sample thickness is
about one-third of the gasket thickness to avoid damage of crystal
and to maintain a good hydrostatic condition. Silicone oil was used
as pressure transmitting media. The pressure calibration was carried
out with ruby fluorescence method \cite{Mao:1986}. High pressure
Raman spectra were collected using inVia Renishaw Raman spectrometer
with a 532 nm exciting laser and 2400 l/mm grating.

For HPXRD measurements, same DAC was used with Ne gas as a pressure-transmitting medium. Pressure was calibrated with equation of state of gold and platinum, which were loaded along with the sample inside the DAC. The experiments were carried out at Advanced Photon Source, GSECARS, 13IDD beamline with wavelength $\lambda$= 0.3344 {\AA}. The 2D diffraction images were integrated with FIT2D software and obtained intensity-2$\theta$ patterns were analyzed using GSAS software to receive the information about the phases and lattice parameters.

All first-principles density functional theory (DFT)
\cite{Hohenberg:1964,Kohn:1965} based simulations have been
performed using the plane-wave based \textit{Vienna ab-initio
simulation package} (VASP)
\cite{Kresse:1993,Kresse:1996,Kresse:1996a}. We have used
projector-augmented wave (PAW) potentials
\cite{Blochl:1994,Kresse:1999}, and the wave functions were expanded
in the plane-wave basis with a kinetic energy cut-off of 350 eV. The
exchange-correlation functional was chosen to be
Perdew-Burke-Ernzerhof (PBE) \cite{Perdew:1996} implementation of
the generalized gradient approximation. Total energies were
converged to less than 10$^{-6}$ eV. In case of structural
relaxation, the positions of the ions were relaxed towards the
equilibrium using conjugate gradient algorithm, till the
Hellman-Feynman forces became less that 0.001 eV/{\AA}. The phonon
band structure has been calculated using the finite difference
method \cite{Kresse:1995,Parlinski:1997}, as realized in the PHONOPY
code \cite{Togo:2015,Togo:2008} in conjugation with DFT as executed
in the VASP code. To cross-check our results, we have calculated the
zone center phonon frequencies using density functional perturbation
theory (DFPT) \cite{Baroni:2001} as implemented in the VASP code.\\

\begin{figure}
\includegraphics[width=0.5\textwidth]{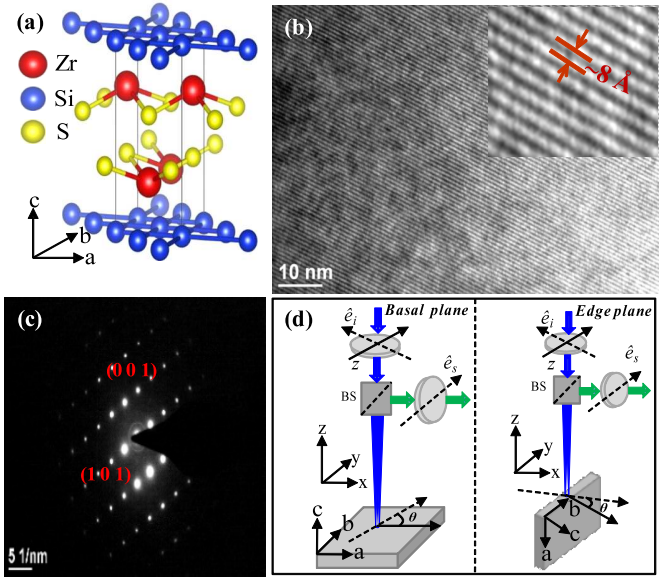}
\caption{(a) Crystal structure of ZrSiS. (b) HRTEM image along the \textit{ac}-plane. Inset shows the magnified image along with the inter-layer
spacing. (c) Selected area electron diffraction (SAED) pattern
obtained through HRTEM measurement. (d) Schematic representation of
the Raman measurement for \textit{basal plane} and \textit{edge
plane} configurations.}
\end{figure}

\section{Results and discussion}

ZrSiS crystallizes in the tetragonal structure with lattice symmetry
$P4/nmm$ (space group no. 129) and point group $D_{4h}$
\cite{Haneveld:1964,Tremel:1987}. Fig. 1(a) shows the crystal
structure of ZrSiS, where \textit{a}, \textit{b} and \textit{c} are
crystallographic axes. Layers of Zr and S are sandwiched between the
Si square nets located in the \textit{ab}-plane. Neighboring S atoms
reside between two Zr layers. The high-resolution transmission
electron microscopy (HRTEM) image of a typical ZrSiS single crystal
is shown along \textit{ac}-plane in Fig. 1(b). High quality
crystalline nature of the samples can be clearly seen with an
inter-layer spacing $\sim$8 {\AA} [shown in the inset of Fig. 1(b)].
The electron diffraction pattern of the crystal is shown in Fig.
1(c) with the Miller indices of the corresponding lattice planes.
Additional characterization details can be found in our earlier
report \cite{Ratnadwip:2017}. Fig. 1(d) schematically illustrates
the experimental set-ups for \textit{basal plane} and \textit{edge
plane} configurations. Here, $\theta$ is the angle of rotation for
the polarization vector and has been discussed later.

\begin{figure*}
\includegraphics[width=0.8\textwidth]{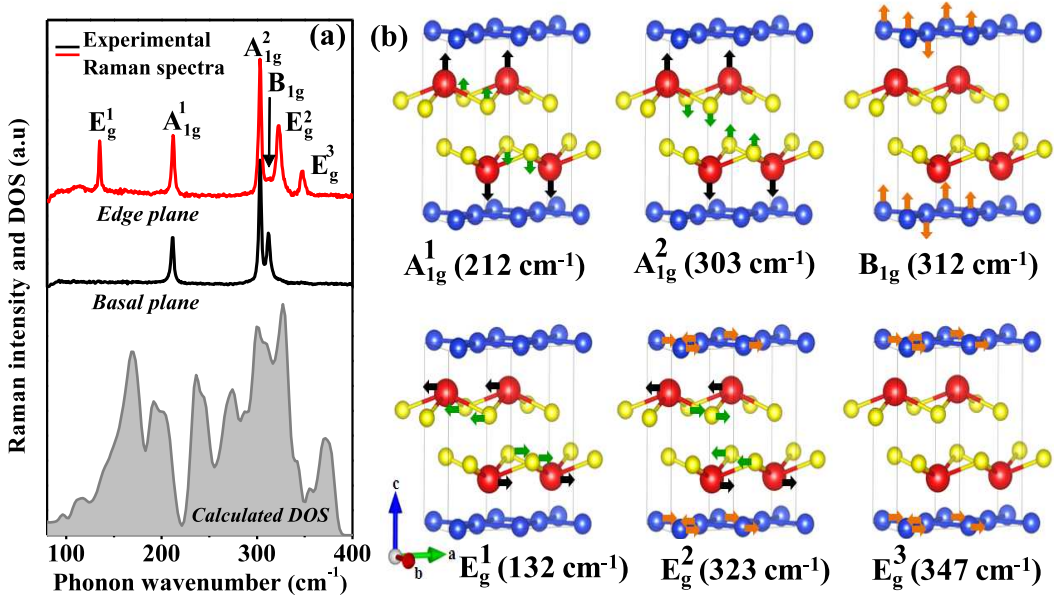}
\caption{(a) Room temperature Raman spectra for \textit{basal plane} (black curve) and \textit{edge plane} (red curve) configurations. The shaded region in the lower panel represents the calculated phonon density of states for ZrSiS. (b) Vibration patters corresponding to the Raman active modes.}
\end{figure*}

From the group symmetry analysis, the zone center optical phonon
modes of ZrSiS can be expressed as
$\Gamma_{ZrSiS}=2E_{u}+2A_{1g}+2A_{2u}+3E_{g}+B_{1g}$
\cite{Salmankurt:2016}. Among them, $A_{2u}$ and $E_{u}$ modes are
IR active, whereas the other 6 modes ($2A_{1g}$, $B_{1g}$ and
$3E_{g}$) are Raman active. In Fig. 2(a), the results of Raman
measurements at room temperature for the two configurations of the
ZrSiS crystal are presented along with the calculated phonon density
of states (PDOS), shown by the filled area. In the \textit{basal
plane} configuration, the incident laser beam excites the
\textit{ab}-plane of the crystal and shows three Raman active modes.
On the other hand, for the \textit{edge plane} measurements, the
crystal is rotated by 90$^{\circ}$ around \textit{b}-axis so that
the laser now falls on the \textit{bc}-plane of the crystal. In this
configuration, all the six Raman active modes have been observed. To
identify the observed modes, we have calculated phonon dispersion
for ZrSiS (Fig. 10 in Appendix) and estimated all
Raman active frequencies at the Brillouin zone center using
first-principles calculations. As shown in Table I, the calculated
frequencies are in excellent agreement with the experimental
results. Note that the $E_{g}$ modes are absent in the \textit{basal
plane} measurements, which can be explained from the lattice
symmetry analysis, discussed later. In Fig. 2(b), the vibrational
patterns of all six Raman active modes have been illustrated. The
$A_{1g}$ optical modes at 212 and 303 cm$^{-1}$ belong to the
anti-symmetric vibration of Zr and S atoms along $c$-axis. The
vibration of Si atoms along $c$-axis with frequency 312 cm$^{-1}$
corresponds to $B_{1g}$. The $E_{g}$ modes at 132, 323 and 347
cm$^{-1}$ come from the in plane vibrations of Zr, Si and S atoms.

\begin{table}
\begin{center}
\caption{Comparison of the calculated and experimental Raman active phonon frequencies.}
\begin{tabular}{|c|c|c|}\hline
Mode &Calc. Freq. &Expt. Freq. \\
 &(cm$^{-1}$)&(cm$^{-1}$)\\ \cline{1-3}
$A^1_{1g}$ &212 &210$\pm$0.5\\
$A^2_{1g}$ &303 &303$\pm$0.5 \\
$B_{1g}$ &312 &312$\pm$0.7 \\
$E^1_{g}$ &132 &134$\pm$0.5 \\
$E^2_{g}$ &323 &321$\pm$0.5 \\
$E^3_{g}$ &347 &347$\pm$0.5 \\
\cline{1-3}
\end{tabular}
\end{center}
\end{table}

\begin{figure}
\includegraphics[width=0.5\textwidth]{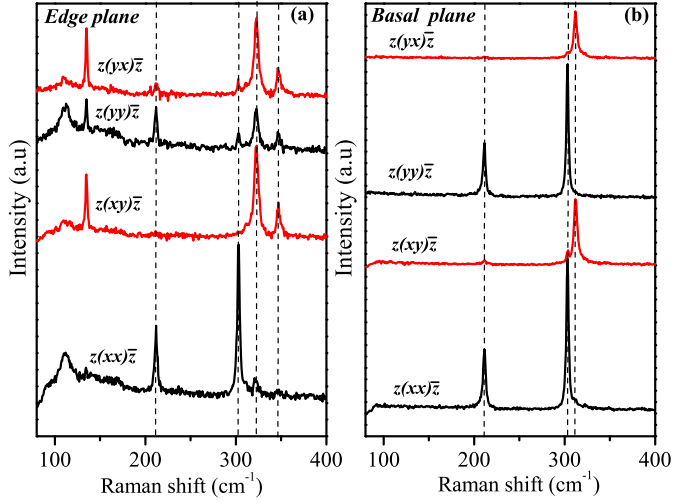}
\caption{Polarized Raman spectra of (a) \textit{edge plane} and (b) \textit{basal plane} configurations.}
\end{figure}

To verify the symmetry of the observed modes, we have measured the
polarized Raman spectra in the backscattering configurations with
light polarization along certain crystallographic directions. We use
Porto's notation for parallel [$z(xx)\bar{z}$ and $z(yy)\bar{z}$]
and perpendicular [$z(xy)\bar{z}$ and $z(yx)\bar{z}$]
configurations. Here, $z$-axis is along the crystallographic $c$-axis
for \textit{basal plane} and antiparallel to $a$-axis for
\textit{edge plane} configuration [see Fig. 1(d)]. The observed
polarized Raman spectra of ZrSiS for both configurations are shown
in Fig. 3. Only $A_{1g}$ modes have been detected when the
measurement was conducted in the parallel channels, i.e.,
$z(xx)\bar{z}$ or $z(yy)\bar{z}$. On the other hand, $B_{1g}$ and
$E_{g}$ modes are present for $z(xy)\bar{z}$ or $z(yx)\bar{z}$
polarization configurations. The observed small intensity of the
forbidden modes at a particular configuration is due to the small
misalignment of the laser polarization with crystallographic axis.

The intensity ($I$) of different Raman active modes obeys a fundamental relation \cite{Yu:2010},
\begin{equation}
I\propto\mid\hat{e_{i}}.\Re.\hat{e_{s}}\mid^{2},
\end{equation}
where $\hat{e_{i}}$ and $\hat{e_{s}}$ are the unit polarization vectors of the incident and the scattered lights, respectively. $\Re$ represents the rank 2 Raman scattering tensor for a particular mode and is determined by the point group symmetry of the material. For ZrSiS, with respect to the principal axes of the crystal, the scattering tensors in the \textit{bc}-plane take the forms,
\[
\Re(A_{1g})=
          \begin{bmatrix}
          \alpha & 0 & 0\\
          0 & \alpha & 0\\
          0 & 0 & \beta
      \end{bmatrix},
\Re(B_{1g})=
      \begin{bmatrix}
     \delta & 0 & 0\\
      0 & -\delta & 0\\
      0 & 0 & 0
      \end{bmatrix},
\]
\[
\Re(E_{g})=
      \begin{bmatrix}
      0 & 0 & -\rho\\
      0 & 0 & \rho\\
      -\rho & \rho & 0
      \end{bmatrix}.\]
Here, $\alpha$, $\beta$, $\delta$ and $\rho$ are Raman scattering components, which can be extracted by fitting the experimental data. When the polarization vectors of the incident and scattered lights are parallel, say along $y$-axis in an orthogonal \textit{xyz} coordinate system, the unit vectors are given by,
\[\hat{e}_{i,xyz}= \begin{bmatrix}
                           0 & 1 & 0
                   \end{bmatrix} ; \hat{e}_{s,xyz}= \begin{bmatrix}

                           0 \\
                           1 \\
                           0
                   \end{bmatrix}.\]
To get the \textit{basal plane} configuration, one has to rotate the crystal by 90$^{\circ}$ around $b$-axis. This operation introduces a rotational matrix, which modifies the scattering tensors in the following forms.
\[
\Re(A_{1g})(basal)=
      \begin{bmatrix}
         \beta & 0 & 0\\
          0 & \alpha & 0\\
          0 & 0 & \alpha
      \end{bmatrix},
\Re(B_{1g})(basal)=
      \begin{bmatrix}
      0 & 0 & 0\\
      0 & -\delta & 0\\
      0 & 0 & \delta
      \end{bmatrix},
\]
\[
\Re(E_{g})(basal)=
      \begin{bmatrix}
      0 & \rho & \rho\\
      \rho & 0 & 0\\
      -\rho & 0 & 0
      \end{bmatrix}.\]
Using these matrices in Eq. (1), one can find that $A_{1g}$
and $B_{1g}$ modes should be present in \textit{basal plane}
configuration, while the intensity of the  $E_{g}$ modes is always zero.
These calculated results from crystallographic symmetries are
consistent with our experimental observations, as shown in Fig.
2(a).

Next, we have studied the complete crystal orientation dependent
Raman spectrum for ZrSiS. We chose the \textit{edge plane}
configuration, as all the modes can be identified in this set up and
rotated the polarization vector in the crystallographic
\textit{bc}-plane (Fig. 11 in Appendix). The intensity
variation of different modes (Fig. 12 in Appendix) can be fitted well using the following equations.
\begin{equation}
I(A_{1g})\propto\mid
\alpha\cos^{2}\theta+\beta\sin^{2}\theta\mid^{2}=C_{1}+C_{2}\cos2\theta+C_{3}\cos4\theta
\end{equation}
\begin{equation}
I(B_{1g})\propto\mid-\delta\cos^{2}\theta\mid^{2}=C_{4}+C_{5}\cos2\theta+C_{6}\cos4\theta
\end{equation}
\begin{equation}
I(E_{g})\propto\mid2\rho\sin\theta\cos\theta\mid^2=C_{7}-C_{8}\cos4\theta
\end{equation}
These relations have been derived from Eq. (1), using the
scattering matrices for the \textit{edge plane} (Details are
provided in Appendix). Here, the parameters $C_{i}$
($i$=1,2,..8) are the functions of the components of Raman
scattering tensors and $\theta$ is the angle between the
polarization vector and crystallographic \textit{b}-axis. All the
modes show four-fold symmetry patterns.

We have also performed the wavelength dependent Raman measurements
for both \textit{basal plane} and \textit{edge plane}
configurations. In Fig. 4, the result for \textit{basal plane}
configuration is shown as a representative. A definite dependency of
the Raman scattering intensity on the energy of the incident laser
is observed. The integrated intensity of the most intense peak
($A^2_{1g}$) shows the highest value at low excitation energy
($\lambda$=785 nm, $E$=1.58 eV) as displayed in the inset of Fig. 4.
The enhancement is due to the coincidence of the laser energy with
an intrinsic electronic transition of ZrSiS as observed in the
reflectivity spectra \cite{Schilling:2017}. In topological
semimetals, the complex interplay of electronic and phonon degrees
of freedom results in a quasielastic (QE) scattering
\cite{Sharafeev:2017}. In the excitation energy dependent Raman
spectra, in contrast to the narrow phonon features, we have observed
a broad and well-defined peak rather than a flat background close to
the laser energy (Yellow shaded area in Fig. 4). This peak can be
attributed to the QE electronic scattering. The evidence for such
scattering in Dirac semimetal Cd$_3$As$_2$ has been reported
recently \cite{Sharafeev:2017}. The line shape of the QE electronic
scattering can be fitted well with the Lorentzian function. In
semimetals, such Lorentzian shape of QE scattering appears due to
the electronic energy density fluctuations \cite{Bairamov:1993}. The
QE scattering peak intensity is highly photon energy dependent
(inset of Fig. 4). This behavior can be understood as a resonant
enhancement and it occurs, when the energy of the scattered photon
matches with one of the excitonic transition energies
\cite{Gillet:2013, Klein:1983}. The occurrence of resonant enhancement of the
phonon and QE scattering at same excitation energy implies that both
the processes involve same intermediate states. A peak with Gaussian
line shape (orange shaded area in Fig. 4) has been observed at 703
cm$^{-1}$, which may be a direct evidence for the single particle or
collective excitations (plasmon-like) \cite{Sharafeev:2017}.

\begin{figure}
\includegraphics[width=0.45\textwidth]{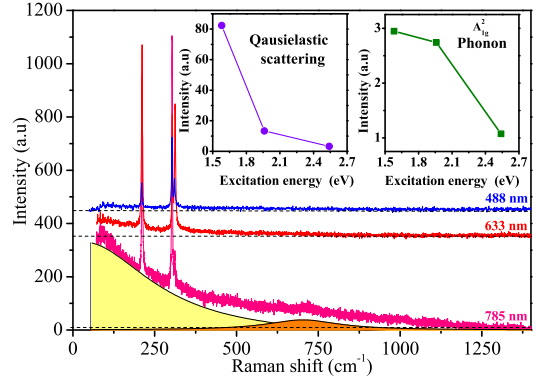}
\caption{Raman spectra of ZrSiS with different incident laser
excitations in \textit{basal plane} configuration. The black dashed
lines correspond to the background. The yellow shaded area
represents the fitting of the QE scattering using Lorentzian line
shape and the orange shaded Gaussian peak is the plasmon maximum at
703 cm$^{-1}$ for 785 nm laser. Insets show the integrated intensity
of the QE and phonon scattering ($A^{2}_{1g}$) as function of
excitation energy.}
\end{figure}

\begin{figure}
\includegraphics[width=0.5\textwidth]{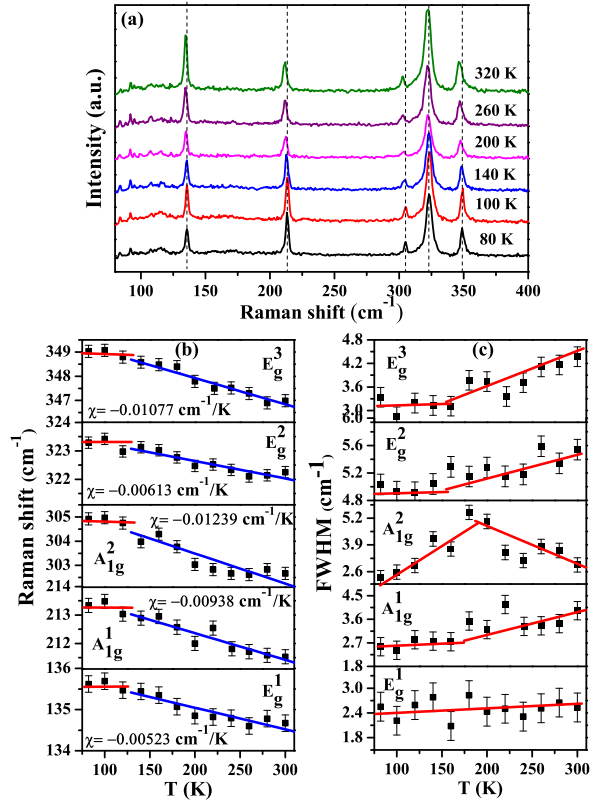}
\caption{(a) Raman spectra of ZrSiS at different representative
temperatures. (b) Raman frequency and (c) FWHM of characteristic
Raman modes as a function of temperature. The solid lines are drawn
as guide to eyes.}
\end{figure}

The temperature dependent Raman spectra for ZrSiS are presented in
Fig. 5(a). The spectra show changes in frequency and full-width at
half-maximum (FWHM) of the modes with an increase in temperature
from 80 K to 320 K. We have fitted all the Raman spectra using
Lorenzian functions to estimate the frequency shift as well as the
line-width of the observed Raman modes in the studied temperature
range. Fig. 5(b) and 5(c) show the temperature-dependent behaviors
of frequency and line-width of the observed Raman modes except for
$B_{1g}$, as the intensity of this mode is very weak. Variations of
peak positions and FWHMs with temperature show two different regions
roughly below and above 150 K. The Raman frequencies of the modes
remain almost same at low temperature region but monotonically red
shifted with the rise of temperature. We know that the factors,
which influence the temperature dependence of Raman shifts, are
phonon-phonon interaction and thermal expansion. Thermal expansion
of the material leads to decrease in frequency of the modes. The
temperature dependence of \textit{i}th phonon energy can be written
as \cite{Eiter:2014}
\begin{equation}
\omega_{ph,i}(T)=\omega_{i}(0)+\Delta_{i}^{1}(T)+\Delta_{i}^{2}(T),
\end{equation}
where $\Delta_{i}^{1}(T)$ is due to thermal expansion and
$\Delta_{i}^{2}(T)$ is related to the phonon-phonon interaction.
$\omega_{i}(0)$ is the phonon frequency at zero temperature, which
is obtained by extrapolating the temperature dependent frequency
change. As $\omega_{i}(0)$ is higher than $\Gamma_{i}(0)$ (FWHM at
zero temperature), the contribution of phonon-phonon coupling in
ZrSiS must be small compared to the shift due to thermal expansion
\cite{Eiter:2014}. To analyze the red shift of the modes at higher
temperature region, we have used a linear dependency
$\omega_{i}(T)=\omega_{i}(0)+\chi_{i}T$ [blue line in Fig. 5(b)],
where $\chi_{i}$ is the first-order temperature coefficient
\cite{Duzynska:2014,Taube:2014}. Using $\chi_{i}$ values, we have
estimated the thermal expansion coefficient and isobaric
mode-Gr\"{u}neisen parameter ($\gamma_{iP}$) for all the modes,
which are shown in Table II. The FWHM of the peaks show two
different slopes with temperature. Except for $A_{1g}^{2}$, all
other modes show a constant FWHM at low temperature and an abrupt
change near 150 K, after which it increases with increasing
temperature. In a perfect crystal, the line-width of the phonon
($\Gamma$) is governed by its interaction with other elementary
excitations as $\Gamma=\Gamma^{an}+\Gamma^{EPC}$. $\Gamma^{an}$ is
the phonon-phonon coupling term, which describes the anharmonic
coupling between phonons and $\Gamma^{EPC}$ represents the
electron-phonon coupling \cite{Ferrari:2007}. $\Gamma^{an}$ is
always present in a system, but $\Gamma^{EPC}$ comes into play only
when the electronic band gap is zero. Here, the FWHM values at lower
temperatures indicate that the contribution from thermal
anharmonicity is lower than the spectrometer resolution (1
cm$^{-1}$). The increase in FWHM with temperature may be a sign of
increased electron-phonon interaction and thus, an increase in
phonon energy dissipation \cite{Bae:2006}. This would lead to
increased electron scattering rate and hence a change in the nature
of temperature dependence of resistivity around 150 K, as observed
in the transport measurements \cite{Ratnadwip:2017}. We can also
calculate the Fermi velocity ($v_{F}$) of the electrons using the
relation \cite{Ferrari:2007},
\begin{equation}
v_{F}=\frac{2Slope(\Gamma)\omega_{\Gamma}}{\Gamma^{EPC}},
\end{equation}
where $Slope(\Gamma)$ is the slope of phonon dispersion curve,
obtained from the quadratic fit to the data presented in Fig. 10 in Appendix and $\omega_{\Gamma}$ is the measured
phonon frequency at zone center. $\Gamma^{EPC}$ values are taken
after deducing maximum possible anharmonicity contribution
(spectrometer resolution 1 cm$^{-1}$) from the measured FWHM at low
temperature. The estimated $v_{F}$ for all the modes (see Table II)
are in excellent agreement with that calculated from quantum
oscillation measurements ($\sim$10$^{5}$ m/s) \cite{Ratnadwip:2017}.
From the wavelength and temperature dependent Raman studies, it is
evident that the lattice dynamics in ZrSiS possesses a strong
correlation with the changes in the electronic properties.

\begin{table}
\begin{center}
\caption{Estimated thermal expansion coefficient($\alpha$),
isobaric and isothermal mode-Gr\"{u}neisen parameters and Fermi
velocity from the measured Raman spectra.}
\begin{tabular}{|c|c|c|c|c|c|}\hline
$\omega_{ph}$&$a_{i}$&$\alpha$&$\gamma_{iP}$&$\gamma_{iT}$& $v_{F}$\\
cm$^{-1}$&(cm$^{-1}$/GPa)&(10$^{-5}$
K$^{-1})$& & &(10$^5$ m/s)\\\cline{1-6}
134 &1.44& 1.39&2.77 &1.23  &1.1  \\
 && (upto 15 GPa)  & & &\\
210 &2.71& 1.48&2.96 &1.88  & 0.9 \\
303 &2.04&1.43 &2.84 & 0.97 & 1.7 \\
321 &3.36&0.98 &1.94 &1.51  &  0.6 \\
347 &4.99&1.20 &2.48 &2.07  &  2.6
\\\cline{1-6}
\end{tabular}
\end{center}
\end{table}

\begin{figure*}
\includegraphics[width=0.7\textwidth]{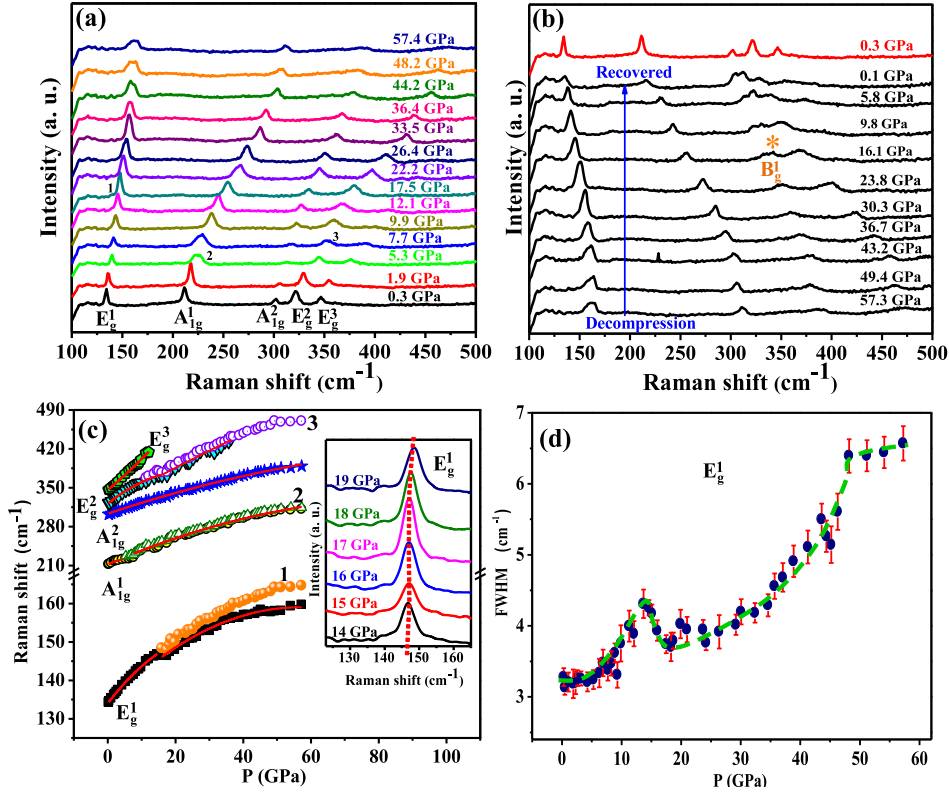}
\caption{Variation of Raman modes with pressure during (a)
compression and (b) decompression up to 57 GPa. (c) Pressure
dependencies of parent phase Raman modes along with new modes. Inset
shows the softening of $E_g^1$ mode at around $\sim$17 GPa. (d)
Variation of FWHM for $E_g^1$ mode as a function of pressure during
compression. The dashed green line is drawn as guide to eyes.}
\end{figure*}

Pressure dependent Raman spectra for ZrSiS have been recorded inside
a diamond-anvil-cell (DAC) in the \textit{edge plane} configuration
for the compression as well as decompression cycles up to 57 GPa as
shown in Fig. 6(a) and 6(b), respectively. During compression, all
the predominant Raman modes other than $E_{g}^{1}$, have been observed to shift towards
higher frequencies with increasing pressure as depicted in Fig.
6(c). $E_{g}^{1}$ mode shows a softening ($\sim$1 cm$^{-1}$) at around $\sim$17 GPa and then starts to harden monotonically with
pressure [Fig. 6(c) inset]. At $\sim$17 GPa, a new band, labeled \textquoteleft1\textquoteright\ emerges at $\sim$ 148 cm$^{-1}$. Another two new bands, labeled \textquoteleft2\textquoteright\ and \textquoteleft3\textquoteright\ appear by splitting of
$A_{1g}^{1}$ and $E_{g}^{2}$ at $\sim$5 GPa and $\sim$10 GPa,
respectively. The bands \textquoteleft2\textquoteright\ and
\textquoteleft3\textquoteright\ are almost inseparable from the
parent modes ($A_{1g}^{1}$ and $E_{g}^{2}$) up to the highest applied pressure
during compression. For all the new bands, with increasing pressure, the frequency increases, intensity reduces and pressure induced broadening
has been observed. All these observed changes are almost reversible
during decompression [Fig. 6(b)]. However, $E_{g}^{2}$ and
$E_{g}^{3}$ modes show large pressure-induced broadening and are not
fully reversible in nature. The existence of all the parent phase
modes along with the new modes and their smooth dependencies with
high pressure, suggest the coexistence of the new phase with the
parent tetragonal phase. It is possible that the new high-pressure
phases have been resulted from the lattice distortion and stacking faults, which occur due
to the destabilization of the weakly bonded S-bilayers as reported in
layered PbFCl \cite{Sorb:2013} and BaFCl systems
\cite{Subramanian:1998}. During pressure release, a small peak,
marked as \textquoteleft$\ast$\textquoteright\ in Fig. 6(b), emerges at $\sim$16 GPa. By
comparing the peak positions at ambient conditions, we have concluded
that it may be the signature of $B_{1g}$ mode, which is not
clearly visible during compression. The pressure dependencies of the
high frequency modes ($E_{g}^{2}$ and $E_{g}^{3}$) are almost linear. For
other modes, it can be fitted with quadratic function and the fitted
curves are shown as solid lines in Fig. 6(c). $E_{g}^{1}$ mode
increases monotonically with pressure after softening, whereas the
mode \textquoteleft1\textquoteright\ starts to separate faster above $\sim$31 GPa. The
pressure derivative and the isothermal mode-Gr\"{u}neisen parameter
($\gamma_{iT}$) have been calculated for all modes using
theoretically estimated bulk-modulus value \cite{Salmankurt:2016} as
shown in Table II. The mismatch between $\gamma_{iP}$ and
$\gamma_{iT}$ may be due to the electron-phonon coupling effect,
which was not excluded for the calculation of $\gamma_{iP}$. We have calculated FWHM for $E_{g}^{1}$ mode and observed an interesting
behavior as shown in Fig. 6(d). FWHM for $E_{g}^{1}$ is maximum at
$\sim$15 GPa and decreases on either side asymmetrically, followed
by a steep rise for higher pressures above $\sim$36 GPa. At around this
pressure range, we have noticed that (i) a new mode
\textquoteleft1\textquoteright\ starts to appear (at $\sim$17 GPa) and (ii) the
softening of $E_{g}^{1}$ occurs.

To correlate our results, let us discuss the geometric structure of
tetragonal ZrSiS, isopuntal with PbFCl \cite{Hughbanks:1995,
Schmelczer} and built up by stacking five 4$^{4}$ square nets into
layers of [..Si$_{2}$-Zr-S-S-Zr-Si$_{2}$..] along the fourfold axis.
Zr atoms form a part of the double-layer of S atoms and build up
sheets of square pyramids [Zr$_{5/5}$S]$_{n}$, which alternate along
\textit{c}-axis with planer Si$_{2n}$ layers \cite{Hughbanks:1995,Schmelczer}. Si atoms are packed into layers
that are twice denser than S layers, resulting appreciable Si-Si
bonding and modest S-S bonding into the unit cell structure. All the
$E_g$ modes correspond to the vibrations, which are confined in \textit{ab}-plane. As our first-principles calculations show, the lowest frequency $E_{g}^{1}$ mode corresponds to relative
motion of the two adjacent S-layers with each Zr layer moving in-phase with its nearest-neighbor S layer, i.e., lacks \textit{c}-axis vibrations. On the other hand, the $A_{1g}^{1}$ mode corresponds to the asymmetric vibrations related to Zr and S atoms and describes the in-phase motion of S-bilayers with Zr-layers on either side. The calculations on phonon density of states also predict that the contribution to the highest optical phonon branches, which correspond to the lowest frequency Raman active modes, comes only from
Si and S atoms. Hence, it is expected that
$E_{g}^{1}$ and $A_{1g}^{1}$ modes (related to S atoms), which
describe the relative motion between the two weakly bonded Si$_{2}$-Zr-S units, can be quite
sensitive to inter-layer interactions under any external
perturbations such as high pressure. The S-bilayers, involving the
stabilizations of the inter-layer arrangements, contribute significantly to
$E_{g}^{1}$ and $A_{1g}^{1}$ modes in \textit{ab}-plane. We also found that under high pressure, low-frequency $E_{g}^{1}$ mode
is more susceptible to change than any other Raman active modes.

The sub-linear pressure dependency of Raman modes indicates that the coupling between the layers
increases under external pressure [Fig. 6(d)]. In contrast to PbFCl, the non-linearity in ZrSiS exits up to the highest applied pressure 57 GPa, suggesting a more rigid structure. This is also supported by their differences in the bulk
modulii, 51 and 144 GPa for PbFCl and ZrSiS, respectively. The
decrease of FWHM followed by the softening of $E_{g}^{1}$ mode at
$\sim$17 GPa is a clear indication of change in the inter-layer
bonding nature, i.e., from layer-type to more isotropic structure. Similar kind of
asymmetric decrease in FWHM for a low-frequency phonon mode was reported
earlier in a topological insulator system Sb$_2$Se$_3$
\cite{Bera:2013}. The authors inferred the phenomenon as pressure
induced electronic topological transition (ETT) from band to topological insulating state. We
found that multiple electronic bands are present at the zone center
of ZrSiS with energy separations comparable to excitations of laser
energies to the inter-band transitions. Some topological insulators
like BiTeI and Bi$_2$Se$_3$ \cite{Genz:2014,Edd:2014} also show
similar results. Gradual splitting of the $A_{1g}^{1}$ mode and
the appearance of modes labeled \textquoteleft1\textquoteright, \textquoteleft2\textquoteright, \textquoteleft3\textquoteright\ can be the indication of isostructural symmetry lowering due
to the instability occurring in the S-bilayers.\\

\begin{figure*}
\includegraphics[width=0.7\textwidth]{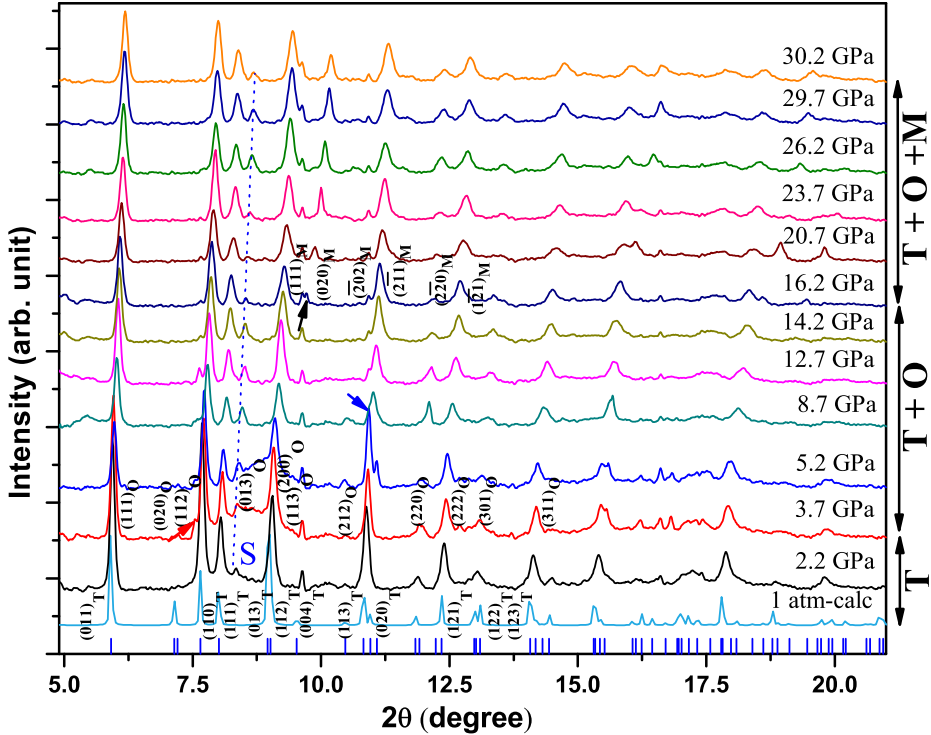}
\caption{The X-ray diffraction patterns with varying pressure. The calculated diffraction pattern with parent tetragonal phase has been plotted to index the high-pressure phases. The slanted arrows (red, blue, and black) show the emergence of new peaks, whereas solid $\Delta$ shows the intensity enhancement of a new peak (020) belonged to the monoclinic phase. `S' is the impurity peak from sulphur and tracked throughout the pressure region (dashed line). The letters `T', `O' and `M' stand for the tetragonal, orthorhombic and monoclinic phases, respectively. The braces on the right side show approximate phase coexistence of the respective phases during compression.}
\end{figure*}

\begin{figure*}
\includegraphics[width=0.6\textwidth]{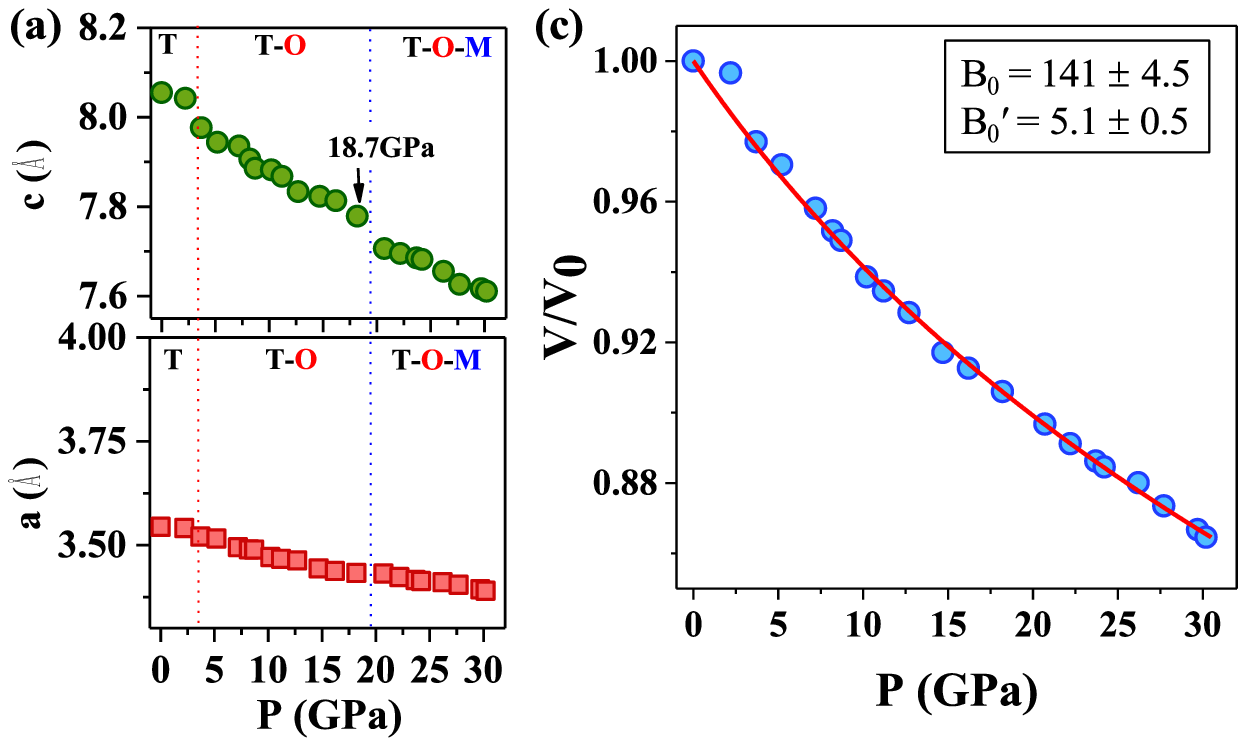}
\caption{(a) Variation of lattice parameter $c$ for the tetragonal phase with pressure. The discontinuity in the same appeared at 18.7 GPa marked with arrow. (b) Variation of lattice parameter $a$ and (c) the normalized volume of the tetragonal phase ($V/V_{0}$) with pressure. The solid line in (c) shows EOS where the values of $B_{0}$=141$\pm$4.5 GPa  and $B_{0}^{\prime}$=5.1$\pm$0.5 are shown. The letters `T', `O' and `M' stand for the tetragonal, orthorhombic and monoclinic phases, respectively.}
\end{figure*}

The synchrotron XRD patterns are recorded as a function of pressure ranging from 2.2 to 30.2 GPa are shown in Fig. 7. A theoretically calculated pattern for 1 atm has been plotted together to index the strongest peaks of the pure tetragonal crystal lattice. At 2.2 GPa, our analysis of HPXRD data assuming the parent $P4/nmm$ crystal structure, resulted in successful indexing of almost all the diffraction peaks in the X-ray pattern other than a peak marked as `S'. We have identified the peak-`S' as an impurity peak of sulphur in our powdered sample and tracked it throughout the measured pressure range (0-30.2 GPa). With the application of pressure, a number of significant changes have been observed, which are marked with arrows in Fig. 7. Some of these changes can be discernible at and below 5 GPa. However, the major parent tetragonal phase is present even at highest pressure 30 GPa in significant amount as supported by our pressure-induced Raman spectra. The extensive studies on the high-pressure structural evolution of the ionic-layered matlokite \cite{Sorb:2013}, alkaline-earth halo-fluorides \cite{Subramanian:1998,Yedukondalu}, metal-oxide hallides \cite{Barninghaausen} and hydride halides composed of sequence of layers, showed the series of symmetry lowering structural transitions from tetragonal to monoclinic phase via the intermediate orthorhombic phase. These results helped us to narrow down the search for the lower symmetry phases in our layered ZrSiS sample. The structure at 3.7 GPa is closely related to tetragonal phase as it can produce all the peaks except the new peaks. Due to large $c$-axis compressibility, the phase must receive a smaller $c/a$ ratio than that at ambient conditions as well as the $a$ or $b$-axis should be longer than the ambient value. Considering these conditions, we have found the presence of an orthorhombic phase (space group: $Pnma$) with lattice parameters $a$=4.23523, $b$=4.9869 and $c$=8.31257 {\AA} at 3.7 GPa, coexisting with the tetragonal lattice $a$=3.52026 and $c$=7.97655 {\AA}. The peaks corresponding to the orthorhombic phase are indexed in Fig. 7, whereas all the other intense peaks are reproduced by the parent $P4/nmm$ structure. The pressure-induced lattice compression and coexistence of phases can be the reasons for the changes observed in the Raman modes ($A_{1g}^{1}$ and $E_{g}^{2}$) at around 5 GPa. A small shoulder peak to the left of (020)$_{T}$ - (212)$_{O}$ (marked with blue arrow in Fig. 7) appeared at $\sim$8.7 GPa. As the pressure is increased, the peak can be clearly identified till 30 GPa. At around 16 GPa, a new peak appeared just right to the (004)$_{T}$ peak. The intensity of this new peak is seen to increase with increasing pressure. All the peaks from the tetragonal and orthorhombic phases are clearly visible, whereas the new peaks can not be indexed with either of these two phases. An analysis at $\sim$16 GPa shows the mixture of three phases; a tetragonal parent phase with coexisting orthogonal and monoclinic phases. Our XRD data analysis shows that the coexistence of the three phases persists up to the highest applied pressure 30 GPa. The newly emerged peak can be indexed as (020)$_{M}$ of monoclinic phase (space group: $P2_{1}/m$) with lattice parameters $a$=7.81417, $b$=3.56305 and $c$=3.56305 {\AA} with $\beta$=107.045$^{\circ}$. A similar type of gradual increase in intensity with increasing pressure has also been observed for BaFBr and BaFCl systems starting from 22 GPa to 60 GPa \cite{Subramanian:1998,Subramanian}. The authors inferred that the phase transition corresponds to the gradual distortion of the tetragonal phase to a monoclinic phase through an intermediate orthorhombic phase. In layered transition-metal dichalcogenides \cite{Bandaru,Gudelli}, the structural phase transition with an increase of intensity for a particular peak can be explained by the changes in Wyckoff's positions of the atoms, corresponding $c/a$ ratio and the variation in chalcogen-metal-chalcogen bond-angles. Our XRD results highly correlate the observed softening of the $E_{g}^{1}$ phonon mode in Raman spectra at around 17 GPa. As we have already discussed, the interlayer arrangements of S-bilayers contribute to the $E_{g}^{1}$ mode in $ab$-plane. We infer that due to high $c$-axis compressibility, there can be a large compression of the electron charge density of the S-ions between the adjacent weakly bonded S-layers along the $c$-axis. Therefore, the anisotropic distribution of such charge density in the $ab$-plane started to distort the lattice during compression. At high pressure above 16 GPa, the tilting of either orthorhombic lattice in the $ab$-plane or tilting of the tetragonal lattice in the $ab$-plane with respect to $c$-axis may cause the orthorhombic to monoclinic phase transition. The effect of such lattice distortions clearly supports our results on pressure-induced Raman spectra. In Fig. 8(a) we have plotted the variation of the lattice parameter $c$ of the parent tetragonal lattice with pressure. The discontinuity of $\sim$0.3 {\AA} is a clear indication of the growing instability in the parent lattice at around $\sim$18.7 GPa. Though the signature of the monoclinic phase appears before 18.7 GPa, it is possible that when a sufficient amount of monoclinic phase sets in, we can observe the measurable discontinuity in the lattice parameters. As shown in Fig. 8(b) and Fig. 8(c), we have not found any discontinuity in lattice parameter $a$ and the calculated normalized unit cell volume ($V/V_{0}$) for the tetragonal phase as a function of pressure. The unit cell volume gradually decreases with pressure without showing any volume collapse. The observed sluggish and smooth nature of the phase transition is due to the small difference in the Gibb's free energy among these three phases \cite{Sorb:2013,Sundarakannan}. Assuming the tetragonal phase for whole pressure range 0-30 GPa, a third order Birch-Murnaghan equation of state (EOS) has been used to fit $V/V_{0}$ with pressure. The obtained bulk-modulus is $B_{0}$=141$\pm$4.5 GPa ($B_{0}^{\prime}$=5.1$\pm$0.5) and agreed very well with the theoretically calculated value 144 GPa \cite{Salmankurt:2016}. The system appears hard to compress with respect to ZrSiTe and ZrSiSe \cite{Salmankurt:2016}.

\section{Conclusions}

In conclusion, we have performed a systematic Raman spectroscopy study on single crystals of ZrSiS. To probe all the Raman active modes, the measurements have been done along two crystallographic planes. The first-principles calculations have been conducted to identify the observed modes as well as the corresponding atomic vibrational patterns. The polarization and crystal rotation dependent Raman studies further verify the mode identification and provide fundamental information about the structural symmetries of the system. The temperature and excitation energy variation measurements reveal resonant enhancement of phonon and quasielastic electronic scatterings, which are the signatures of complex electron-phonon interaction in ZrSiS. Furthermore, the high-pressure Raman spectra show vibrational anomalies and the appearance of new modes, suggesting possible structural transitions. From the structural analysis, we conclude that such modification in crystal structure may also lead to electronic topological transition, as observed in several topological materials. Our high-pressure synchrotron x-ray diffraction measurements show very interesting pressure-induced structural transition from tetragonal to monoclinic phase via an intermediate orthorhombic phase with the absence of any volume collapse. The three phases coexist up to the highest applied pressure (30 GPa) and the structural transformations agree well with the high-pressure Raman spectroscopy results. Therefore, this report present a detailed study on the lattice dynamics and electron-phonon interaction in ZrSiS and may also provide the ground for subsequent investigations in the present system as well as other iso-structural compounds.

\textit{Note added:} While analyzing our data, we came across a report by Zhou \textit{et al.} \cite{Zhou:2017} on the Raman study of ZrSiS under ambient conditions. Contrary the report, here we have performed a detailed polarization-resolved and crystal-angle-resolved Raman measurements along with temperature and pressure dependent studies.\\

\begin{center}
\textbf{APPENDIX}\\
\end{center}

\begin{center}
\textbf{A. HIGH-PRESSURE MEASUREMENTS.}\\
\end{center}

In Fig. 9, we have shown the set-up of diamond-anvil-cell (DAC) for high-pressure measurements.

\begin{figure}
\includegraphics[width=0.4\textwidth]{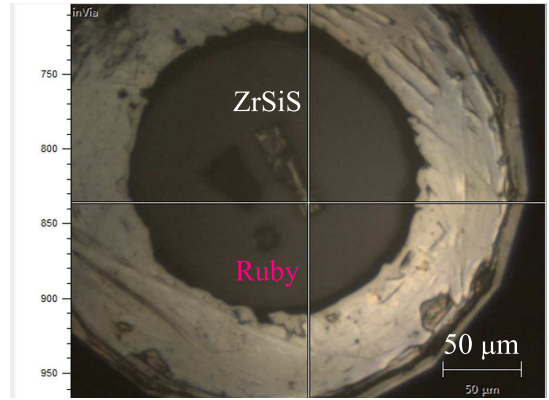}
\caption{ZrSiS single crystal sample and ruby inside DAC for high-pressure experiment.}
\end{figure}

\begin{center}
\textbf{B. CALCULATION OF PHONON DISPERSION.}\\
\end{center}

The theoretically calculated phonon dispersion spectra for ZrSiS are shown in Fig. 10.

\begin{figure}
\includegraphics[width=0.4\textwidth]{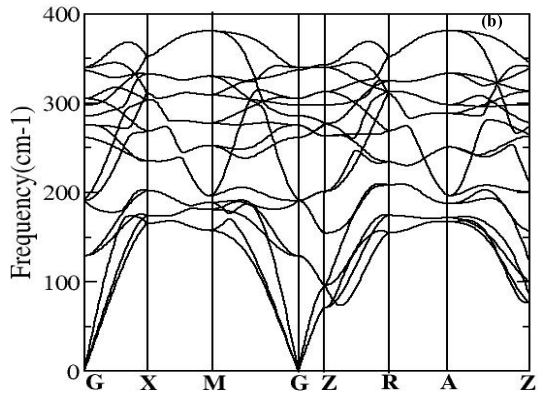}
\caption{Calculated phonon dispersion spectra for ZrSiS.}
\end{figure}

\begin{figure}
\includegraphics[width=0.2\textwidth]{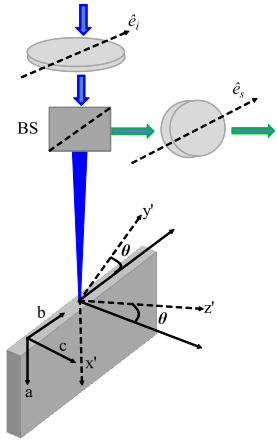}
\caption{Schematic of the experimental set up for crystal orientation dependent Raman spectroscopy. The crystal is rotated around \textit{a}-axis, which is equivalent to rotation of the laser polarization vector in \textit{bc}-plane.}
\end{figure}

\begin{figure*}
\includegraphics[width=0.8\textwidth]{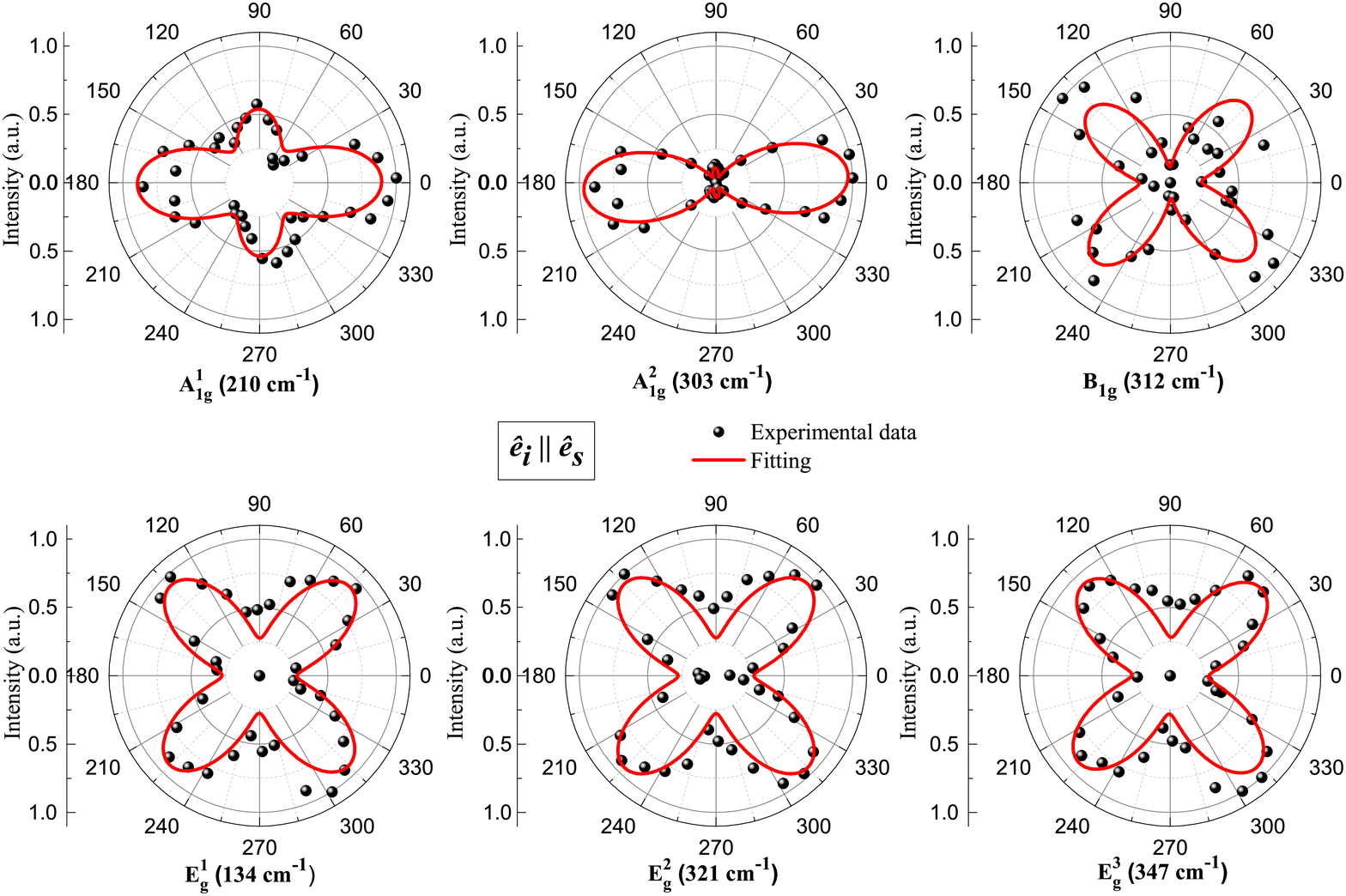}
\caption{Angle dependence of Raman mode intensities in parallel
configuration (\textit{\^{e}$_i\parallel$ \^e$_s$}) on \textit{edge
plane}.}
\end{figure*}

\begin{center}
\textbf{C. CRYSTAL ORIENTATION DEPENDENT RAMAN SPECTROSCOPY.}\\
\end{center}

To describe the Raman spectrum for different crystal orientation and
polarization geometry, let us introduce a new Cartesian coordinate
system, where the x$^\prime$-direction coincide with
crystallographic \textit{a}-axis. Whereas, the y$^\prime$ and
z$^\prime$-directions make an arbitrary angle ($\theta$) with the
crystallographic \textit{b} and \textit{c}-axis, respectively [Fig. 11]. In this new representation, the scattering tensors are
transformed as
\begin{equation}
\Re_{x^{\prime}y^{\prime}z^{\prime}}=\Phi_{x{^\prime}y^{\prime}z^{\prime}}\Re\tilde{\Phi}_{x^{\prime}y^{\prime}z^{\prime}},
\end{equation}
where $\Phi_{x^{\prime}y^{\prime}z^{\prime}}$ is the orthogonal transformation matrix
\[\Phi_{x^{\prime}y^{\prime}z^{\prime}}= \begin{bmatrix}
                        1 & 0 & 0 \\
                        0 & \cos\theta & \sin\theta \\
                        0 & -\sin\theta & \cos\theta
                   \end{bmatrix}\]
and $\tilde{\Phi}_{x^{\prime}y^{\prime}z^{\prime}}$ is its inverse. Using these transformed matrices in Eq. (1) of the main text, we have derived the intensity relations [Eq. (2), (3) and (4) in the main text] for different Raman active modes.

To measure the crystal orientation dependent Raman scattering of ZrSiS,
we chose the \textit{edge plane}, as all  modes can be identified in
this configuration. Initially, we have fixed the polarization
vectors of the incident and scattered lights parallel to each other
and along crystallographic $b$-axis. Note that at this point all the
axes of the Cartesian coordinate system coincide with the
crystallographic axes, i.e., $\theta$=0. Now, we start to rotate the
single crystal in the y$^{\prime}$z$^{\prime}$-plane in small steps
and note the intensity variation of different modes. Although this
method is equivalent to the rotation of the polarization vector
along crystallographic \textit{bc}-plane, rotation of the sample has certain
advantages. To change the polarization direction of both incident
and scattered lights, one has to use additional optical pieces in
the light path and adjust their polarization angles at each step.
Such setup unavoidably introduce experimental uncertainties in the
polarization angle and also can modulate the intensity of different
modes \cite{Ma:2016}. The obtained angular dependence of the
normalized intensity profile is shown in Fig. 12 (scattered points)
for different modes as polar plots. The experimental data have been fitted well using Eq. (2), (3) and (4) (solid red
lines in Fig. 12), which have been derived from the crystal symmetry of ZrSiS.

\end{document}